\documentclass[twocolumn,prb,amsmath,amssymb,amsfonts,superscriptaddress,floatfix,aps,showpacs]{revtex4-1}

\usepackage{graphicx}% Include figure files
\usepackage{dcolumn}% Align table columns on decimal point
\usepackage{bm}% bold math
\usepackage{rotating} % rotate figure
\usepackage{color}
\usepackage{subfigure}
\usepackage{natbib}
\begin{document}
\newcommand{\red}{\color{red}}

\title{The valence and spectral properties of rare-earth clusters}
\date{\today}
\author{L. Peters}
\email{L.Peters@science.ru.nl}
\affiliation{
Radboud University Nijmegen, Institute for Molecules and Materials, NL-6525 AJ Nijmegen, The Netherlands
 }
\author{I. {Di Marco}}
%\email{mattias@physics.uu.se}
\affiliation{
Department of Physics and Astronomy, Uppsala University,
Box 516, SE-75120, Uppsala, Sweden
 }
\author{M.S. {Litsarev}}
%\email{mattias@physics.uu.se}
\affiliation{
Skolkovo Institute of Science and Technology, Novaya St. 100, Skolkovo, Odintsovsky District, 143025 Moscow Region, Russia
 }
\author{A. Delin}
%\email{mattias@physics.uu.se}
\affiliation{
Department of Physics and Astronomy, Uppsala University,
Box 516, SE-75120, Uppsala, Sweden
}
\affiliation{
Department of Nano and Materials Physics, Royal Institute of Technology (KTH), Electrum 229, SE-16440 Kista, Sweden
 }
\author{M.I. Katsnelson}
%\email{mattias@physics.uu.se}
\affiliation{
Radboud University Nijmegen, Institute for Molecules and Materials, NL-6525 AJ Nijmegen, The Netherlands
 }
\author{A. Kirilyuk}
%\email{mattias@physics.uu.se}
\affiliation{
Radboud University Nijmegen, Institute for Molecules and Materials, NL-6525 AJ Nijmegen, The Netherlands
 }
\author{B. Johansson}
%\email{olle.eriksson@physics.uu.se}
\affiliation{
Department of Physics and Astronomy, Uppsala University,
Box 516, SE-75120, Uppsala, Sweden
 }
\affiliation{
Department of Materials Science and Engineering, Royal Institute of Technology (KTH), SE – 100 44 Stockholm, Sweden
 }
\author{B. Sanyal}
%\email{olle.eriksson@physics.uu.se}
\affiliation{
Department of Physics and Astronomy, Uppsala University,
Box 516, SE-75120, Uppsala, Sweden
 }
\author{O. Eriksson}
%\email{olle.eriksson@physics.uu.se}
\affiliation{
Department of Physics and Astronomy, Uppsala University,
Box 516, SE-75120, Uppsala, Sweden
 }

\begin{abstract}
The rare-earths are known to have intriguing changes of the valence, depending on chemical surrounding or geometry. Here we make predictions from theory that combines density functional theory with atomic multiplet-theory, on the transition of valence when transferring from the atomic divalent limit to the trivalent bulk, passing through different sized clusters, of selected rare-earths. We predict that Tm clusters show an abrupt change from pure divalent to pure trivalent at a size of 6 atoms, while Sm and Tb clusters are respectively pure divalent and trivalent up to 8 atoms. Larger Sm clusters are argued to likely make a transition to a mixed valent, or trivalent, configuration. The valence of all rare-earth clusters, as a function of size, is predicted from interpolation of our calculated results. We argue that the here predicted behavior is best analyzed by spectroscopic measurements, and provide theoretical spectra, based on dynamical mean field theory, in the Hubbard-I approximation, to ease experimental analysis.
\end{abstract}
\maketitle
\noindent

\section{Introduction}
Recently the total magnetic and dipole moments of isolated rare-earth clusters (Pr, Tb, Ho and Tm) in the gas phase have been measured experimentally for a size range of 5-30 atoms~\cite{chris1,chris2}. These experiments show a very interesting and unexpected behavior completely different from the bulk. For example, for Tb clusters the magnetic moment oscillates heavily as function of cluster size, while for Tm and Pr clusters there is almost no size dependence. Further, there appears to be a large electric dipole moment for certain Tm cluster sizes. Understanding the principles behind this behavior is important not only from a fundamental point of view, but also for possible applications at nanoscale. 

However, before the problem of magnetism or any problem in general for rare-earth clusters can be addressed, an absolutely crucial foundation has to be layed. Namely the knowledge of the number of 4$f$ electrons or equivalently the number of $spd$ electrons of the rare-earth clusters is required. This is often referred to as the valence. This valence for rare-earth clusters is not trivial, because in general as an isolated atom a rare-earth has one 4$f$ electron more than in its bulk form~\cite{mac1}. 

In order to study magnetic properties of rare-earth systems, knowledge of the number of 4$f$ electrons constituting the local magnetic moments and the number of $spd$ electrons mediating the coupling between them is required. Further, the valence of the rare-earths is known from previous works to depend delicately on chemical surrounding or geometry. For instance, it is known that the surface of elemental Sm is divalent~\cite{surface1}, whereas the bulk is trivalent. Overall, the valence is intimately coupled to many of the important properties of the rare-earths. It is this crucial foundation of the valence that is here addressed for clusters. We evaluate the electronic structure and the total energy of such clusters, using density functional theory (DFT)~\cite{dft1,dft2}, combined with the Born-Haber cycle~\cite{anna2,borje2}. We selected three elements Sm, Tb and Tm, since they are known to have small energy difference between a trivalent and divalent configuration in the bulk~\cite{anna2}, and due to recent experimental interest~\cite{chris1,chris2}. 

It is well known for rare-earth bulk systems that DFT in its conventional localized density approximation (LDA) or generalized gradient approximation (GGA) form is inadequate~\cite{skriv1,min1,duth1,lars1}. This failure is caused by the localized 4$f$ electrons, for which the electron-electron repulsion is strong, which cannot be described properly by functionals derived in the limit of a nearly uniform electron gas. Therefore, we chose to work with the DFT+DMFT~\cite{dmft1} approach in the limit of zero hybridization, i.e. the Hubbard-I approximation (HIA)~\cite{hia1}, is used to calculate the spectral properties of these clusters. Before a HIA calculation can be performed, the geometry, the valence stability, Hubbard $U$ parameter and the first 4$f$ peak position below the Fermi level of the cluster must be known. For the geometry calculations, the 4$f$ electrons are made chemically inert by treating them as part of the core~\cite{petu}. In contrast, to calculate the valence stability and first 4$f$ peak position below the Fermi level, we follow Refs.~\onlinecite{anna2,borje2}. 

The Born-Haber cycle has already been used with great success and accuracy to calculate the valence and first 4$f$ peak below the Fermi level for the whole 4$f$ bulk series~\cite{anna2,borje2}. It is now well established that with the exception of Eu, Yb and the $\alpha$-phase of Ce, all rare-earth elements form trivalent configurations in the solid~\cite{mac1}. Eu and Yb are divalent, because this configuration provides a half-filled or filled 4$f$ shell~\cite{mac1}. On the contrary the isolated rare-earth atoms are all divalent with the exception of La, Ce, Gd and Lu, which are trivalent. The transition from atom to solid can be seen as a function of clusters with increasing size. The valence stability of the end points is known precisely. For Sm, Tb and Tm the atom is divalent whereas the bulk is trivalent. The valence of clusters is completely unknown and it is reasonable to ask how the transition from divalent to trivalent occurs as a function of increasing cluster size, for what size of clusters it happens and if mixed valence configurations are possible.

\section{Theory}
\subsection{Valence}
In case of the valence the aim is to calculate the total energy difference between a divalent, $f^{n+1}[spd]^{2}$, trivalent, $f^{n}[spd]^{3}$ and mixed valence configuration. Unfortunately, these total energies (differences) are not directly accessible for rare-earth systems in conventional DFT (LDA or GGA)~\cite{anna1,anna2}. However, for localized and thus strongly correlated systems the Born-Haber cycle can be used for this purpose. The idea is to exploit the fact that the 4$f$ shell is so localized that it is essentially the same in the atom as in the isovalent cluster. This gives the opportunity to combine DFT calculations, in which intra and inter 4$f$ couplings are neglected, with atomic experimental information to include these couplings.  

\begin{figure}[bt]
\begin{center}
\includegraphics[trim=10 520 5 10, clip, width=8cm]{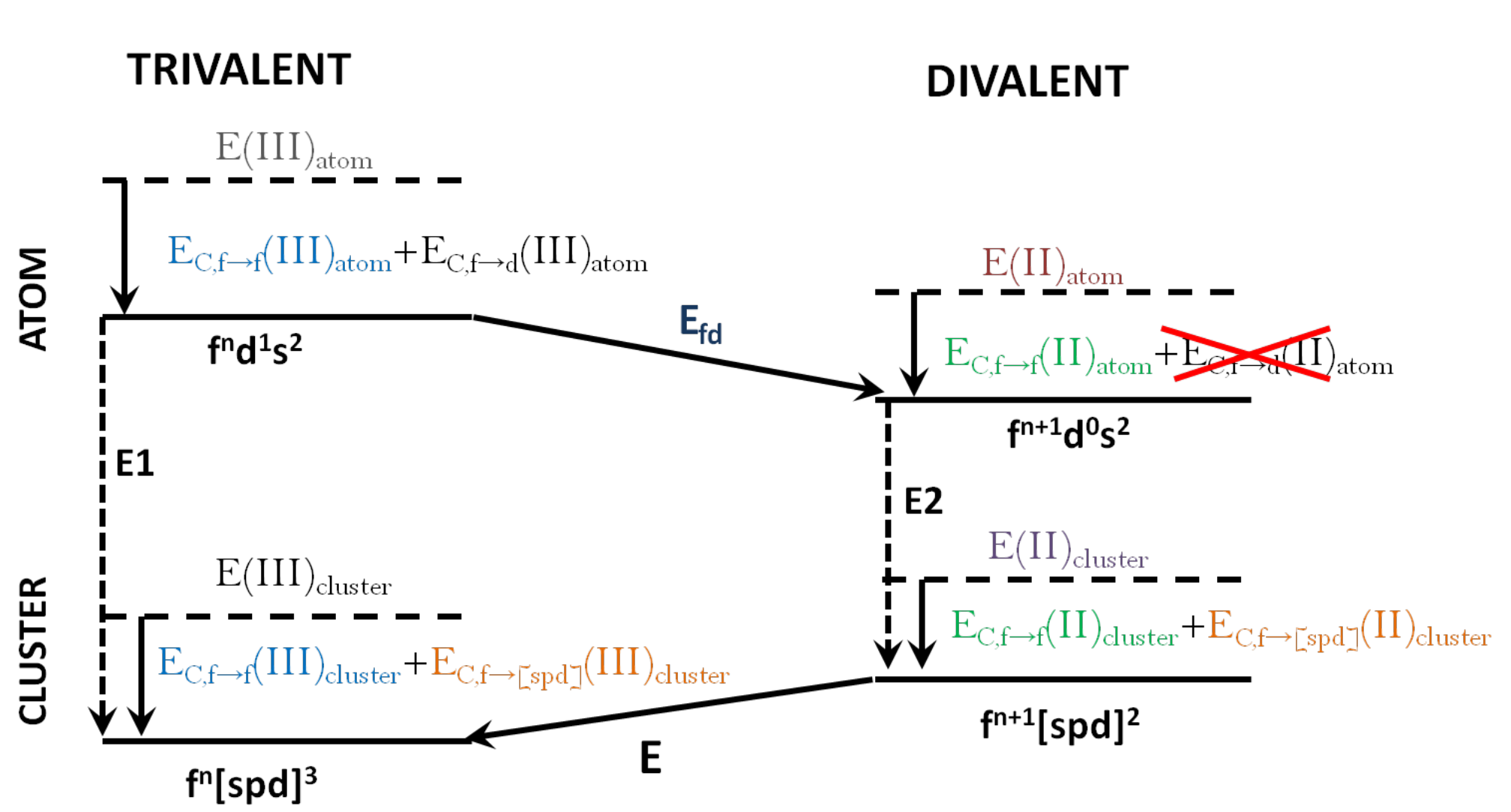}
\end{center}
\caption{Schematical picture of the Born-Haber cycle\cite{anna2}. The dotted lines represent total energies without intra and inter 4$f$ coupling, while the full lines corresponds to full total energy. Further, $E_{fd}$  is the atomic $f$ to $d$ promotion energy, $E_{C,f\to d}(II)_{atom}$ and $E_{C,f\to d}(III)_{atom}$ refer to the coupling between the 4$f$ and 5$d$ shell of respectively the divalent and trivalent atom, $E_{C,f\to[spd]}(II)_{cluster}$ and $E_{C,f\to[spd]}(III)_{cluster}$ correspond to the coupling between the 4$f$ shell and $spd$ states of respectively the divalent and trivalent cluster, and $E_{C,f\to f}$ terms refer to intra 4$f$ shell couplings. } 
\label{BORNHABER}
\end{figure}

In Fig.~\ref{BORNHABER} the Born-Haber cycle is schematically depicted for the computation of the energy difference $E$ between a pure divalent and trivalent configuration. From this figure it is immediately clear that $E$ cannot be computed directly in conventional DFT due to the lack of a proper description of the localized 4$f$ shell. To a good approximation the difference between the inter 4$f$ couplings, $E_{C,f\to[spd]}(III)_{cluster}$ and $E_{C,f\to[spd]}(II)_{cluster}$, can be neglected \cite{anna1,anna2}. Thus it remains to compute the energy difference between the intra 4$f$ couplings, $E_{C,f\to f}(III)_{cluster}$ and $E_{C,f\to f}(II)_{cluster}$. This problem can be circumvented by going around the cycle via the atomic energies, since these intra 4$f$ couplings then cancel with their isovalent atomic counterparts. However, by going around the cycle in this way, three new quantities have to be introduced, i.e. $E_{fd}$, $E_{C,f\to d}(III)_{atom}$, and $E_{C,f\to d}(II)_{atom}$. The former is the energy required to promote a 4$f$ electron to the 5$d$ shell in the atom and the latter two represent the coupling energy between the 4$f$ shell and the 5$d$ shell. Since there is no 5$d$ electron in the divalent atom, $E_{C,f\to d}(II)_{atom}$ is zero, while the other two are known experimentally~\cite{borje1,anna2,atomexp}. 

Besides the introduction of these experimentally known atomic quantities, the energy difference between the decoupled (intra and inter 4$f$ coupling neglected) isovalent atom and cluster should now be accuratly computed. This energy difference is often referred to as generalized cohesive energy and is known to be reproduced very well by DFT~\cite{anna1,anna2}.

The concept above can be easily extended to more general configurations, e.g. mixed valence configurations. The expression for the energy difference per atom $E$ between a pure trivalent and mixed valence configuration thus becomes,
\begin{equation}
\begin{gathered}
E=\bigl[E(III)_{atom}-E(III)_{cluster}\bigr]- \\
\bigl[\frac{n_{div}}{n_{tot}}E(II)_{atom}+\frac{n_{triv}}{n_{tot}}E(III)_{atom}-E(mix)_{cluster}\bigr]-\\
\frac{n_{div}}{n_{tot}}\bigl[E_{fd}+E_{C,f\to d}(III)_{atom}\bigr].
\end{gathered}
\label{BORNMIX}
\end{equation}
\newline
Here $n_{tot}$ is the total number of atoms in the pure trivalent and mixed valence cluster, and $n_{div}$ and $n_{triv}$ correspond respectively to the number of divalent and trivalent atoms in the mixed valence cluster. Further, $E(III)_{cluster}$ and $E(mix)_{cluster}$ correspond to the total energy per atom with intra and inter 4$f$ coupling neglected for a trivalent and mixed valence configuration of a cluster. Similarly $E(III)_{atom}$ and $E(II)_{atom}$ are the total energy with intra and inter 4$f$ coupling neglected of a trivalent and divalent atom. Finally, $E_{fd}$ and $E_{C,f\to d}(III)_{atom}$ are the atomic correction energies that are obtained from experiment. Note that the first term between square brackets in Eq.~\ref{BORNMIX} is the generalized cohesive energy of a trivalent configuration. Furthermore, when $E(mix)_{cluster}$ would correspond to a purely divalent configuration ($n_{div}=n_{tot}$ and $n_{triv}=0$ in Eq.~\ref{BORNMIX}), then the second term between square brackets in Eq.~\ref{BORNMIX} refers to the divalent generalized cohesive energy.

%\subsection{Hubbard $U$ parameter}

\subsection{First 4$f$ peak position}
For the explanation of the calculation of the first 4$f$ peak below the Fermi level, Ref.~\onlinecite{borje2} is followed. In this work the first 4$f$ peak below the Fermi level is calculated for all elemental bulk rare-earth systems. The first thing to consider is that the time scale on which a photo-emission process takes place is too short for the geometry to follow. Second, consider a bulk system from which in one unit cell a 4$f$ electron is removed. This process can be artifically divided into two steps: (1) a 4$f$ electron is promoted to the valence band and (2) an electron is adiabatically taken out of the valence band. In bulk the ionization energy of the ground state and the state with one 4$f$ electron promoted to the valence band are virtually the same due to the screening of this 4$f$ hole by the valence electrons. This means that with respect to this ionization energy the first 4$f$ peak position corresponds to the total energy difference between the ground state and the state in which one atom is replaced with one 4$f$ electron less and one valence electron more. Thus again the Born-Haber cycle can be exploited here. 

For clusters however it is not clear whether the valence electrons are able to fully screen the 4$f$ hole. In other words in case of the bulk there is an infinite number of sites of which only one is changed by the promotion of a 4$f$ electron to the valence band. However, in case of our clusters thereare less than 9 sites, which means that we cannot neglect this change of one site on forehand. Thus, there could be a difference in ionization energy between the ground state and a state in which one 4$f$ electron is promoted to the valence states with respect to the ground state. This difference is estimated by the difference in the eigenvalues of the highest occupied Kohn-Sham orbital of these states.

\section{Details of calculations}
\label{doc}
All the calculations in this report are performed with a full potential linear muffin-tin orbital (FP-LMTO) method~\cite{rsptbook}. A GGA parametrization of the exchange-correlation functional as formulated by Perdew, Burke, and Ernzerhof is used~\cite{gga1}. Since the used FP-LMTO program is originally designed for periodic bulk systems, a large cubic unit cell of 16~{\AA} dimension is used to prevent the interaction between clusters of different unit cells. Further, a calculation for the gamma point only is done. The basic geometrical and basis setup is the same for all calculations described above. A muffin-tin radius of 2.6~a.u. is used. The main valence basis functions are chosen as 6$s$, 6$p$ and 5$d$ states, while 5$s$ and 5$p$ electrons are treated as pseudocore in a second energy set~\cite{rsptbook}. The 4$f$ states are treated as valence states for the Hubbard-I calculations, while for all the other calculations they are treated as core states. In the latter case 5$f$ states are instead added to the valence electrons, in order to have basis functions with $f$ angular character. Three kinetic energy tails were used for 6$s$ and 6$p$ states~\cite{rsptbook}, with values -0.3, -2.8 and -1.6~Ry. Further, a Hubbard $U$ of 8~eV is used for the Hubbard-I calculations of the spectral properties of the Sm, Tb and Tm clusters. This value of $U$ is commonly used for the 4$f$ shell of the rare-earths~\cite{lars1,impl1,leb1,leb2}. In addition to the Hubbard $U$ parameter also the onsite exchange interaction $J$ is needed. For the $J$ it is well known that it is almost system independent~\cite{hubj1,hubj2} and is therefore taken to be 1~eV~\cite{impl1,leb1,leb2}.

Details on the implementation of the Hubbard-I routine, that is used in this work, are given elsewhere~\cite{impl1,impl2,impl3,impl4,impl5}, and we refer the reader to those studies for a complete description of our methods. The double counting is fixed by adjusting the first 4$f$ peak below the Fermi level to the one calculated (see Fig.~\ref{HUBBARDU_4f}) and by the required number of 4$f$ electrons (see Fig.~\ref{VALENCERESULTS}). Namely within the Hubbard-I approximation the local self energy is obtained from the following atomic like problem

\begin{equation}
\hat{H}_{\boldsymbol{R}}^{at}=\hat{H}_{\boldsymbol{R}}^{DFT}+\hat{H}_{soc}+\hat{H}_{U}-\mu_{at}\sum_{\xi} \hat{c}_{\boldsymbol{R}\xi}^{*} \hat{c}_{\boldsymbol{R}\xi}
\tag{S2}
\label{hiaimp2}
\end{equation}
\newline
Here $\hat{c}_{\boldsymbol{R}\xi}^{*}$ and $\hat{c}_{\boldsymbol{R}\xi}$ correspond to the creation and annihilation operators of local strongly correlated state $\lvert \boldsymbol{R},\xi \rangle$. The first term $\hat{H}_{\boldsymbol{R}}^{DFT}$ contains the DFT single particle Hamiltonian projected onto the strongly correlated states. The second term contains the spin-orbit coupling effects and the third term the onsite Coulomb repulsion effects between the strongly correlated orbitals. Finally, the last term contains the chemical potential $\mu_{at}$. This term is used to embed the atom in the cluster. Furthermore, it is used as a double counting correction. Namely some effects of the $\hat{H}_{U}$ term are already included in the $\hat{H}_{\boldsymbol{R}}^{DFT}$ term. There are different possibilities to choose for the double counting~\cite{impl1}. We treat the chemical potential as an adjustable parameter in such a way that it is determined by the number of 4$f$ electrons and the first 4$f$ peak below the Fermi level.  

For the calculation of the valence different starting geometries have been considered and optimized for each valence configuration separately. The wrapped polyhedron method was used to optimize the structures~\cite{simplex}. For the other calculations, i.e. the first 4$f$ peak position below the Fermi level and 4$f$ PDOS, the geometry is fixed to the thus found lowest energy geometry.

\section{Results}
\subsection{Valence stability}
\begin{figure}[bt]
\includegraphics[trim=9 450 15 5, clip, width=8cm]{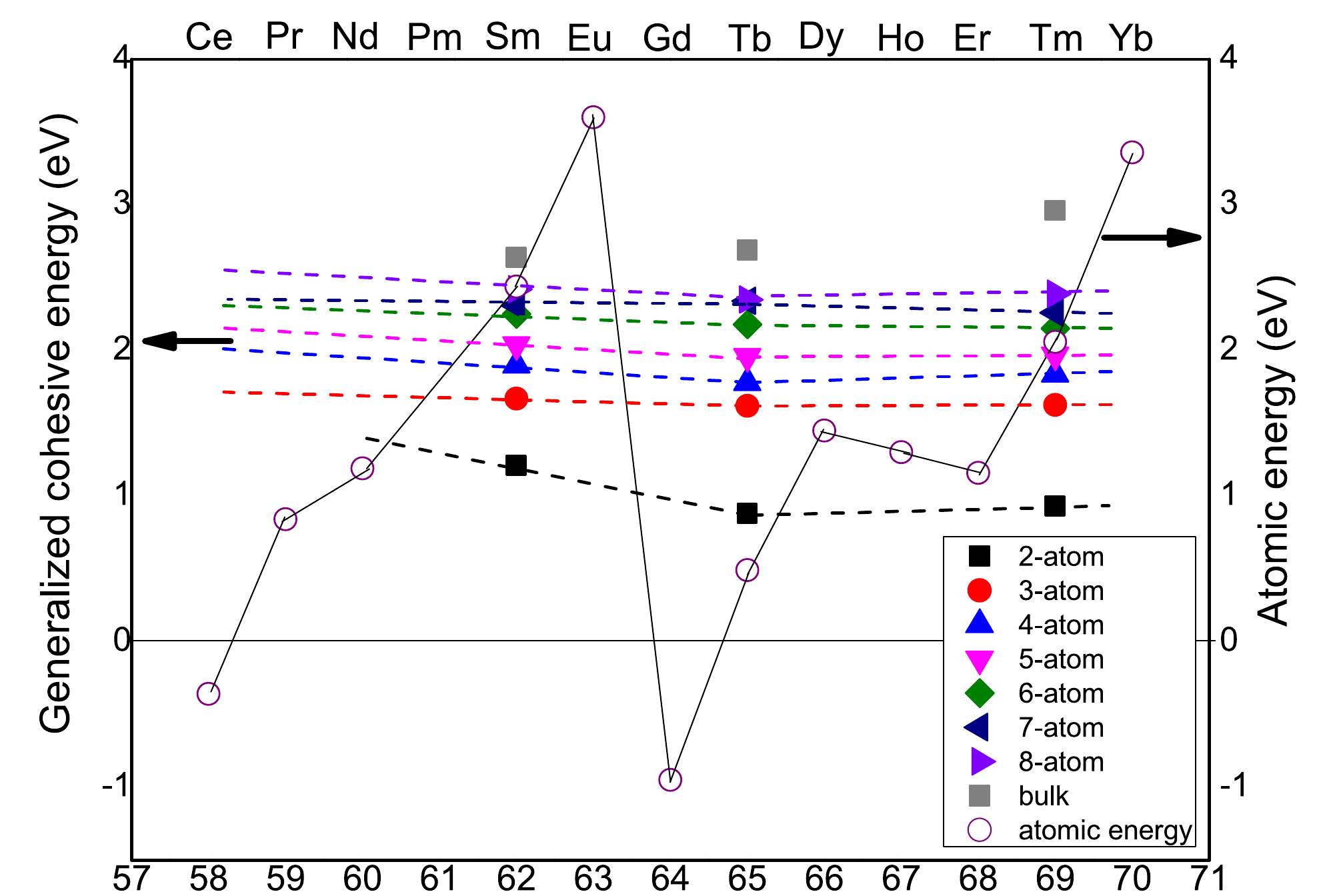}
\caption{Results of valence stability calculation for Sm, Tb and Tm clusters from 2-8 atoms. On the left vertical axis the difference between a pure divalent and trivalent generalized cohesive energy (in eV) is printed for the different cluster sizes. The open circles connected by the solid line correspond to the atomic correction energy (in eV)~\cite{anna2}, shown on the right vertical axis. The dashed lines represent the simple linear extrapolations between the Sm, Tb and Tm data points. The points where the solid line of the atomic correction energy crosses the dashed line correspond to the valence transitions. In the cases where the dashed line is below the solid line the system is divalent and in the opposite cases it is trivalent. Finally, the grey squares correspond to the bulk generalized cohesive energies.} 
\label{VALENCERESULTS}
\end{figure}

By applying the Born-Haber cycle to Sm, Tb and Tm clusters in a size range of 2-8 atoms we obtain the valence (see Fig.~\ref{VALENCERESULTS}). In this graph, we show the difference in the generalized cohesive energy between a pure trivalent and divalent configuration for different sized clusters. We also show the atomic correction energies ($E_{fd}$ plus $E_{C,f\to d}$ as described in Fig.~\ref{BORNHABER}). From Eq.~\ref{BORNMIX} it is clear that when the generalized cohesive energy difference is larger than the sum of atomic correction energies a trivalent state is more favorable. In general, Fig.~\ref{VALENCERESULTS} shows that the generalized cohesive energy becomes larger for clusters with more atoms, since this allows for more chemical bonds to be made. Hence one may expect in general that as the clusters become larger, the divalent state of the atom should become less favorable. This is indeed what Fig.~\ref{VALENCERESULTS} shows. Fig.~\ref{VALENCERESULTS} also shows that Sm and Tb clusters are respectively divalent and trivalent for all clusters investigated here. We find however that for Sm a transition from divalent to trivalent clusters should occur for cluster sizes just above 8. A different behaviour is observed for Tm, since there is a change from divalent to trivalent at a size of 6 atoms. For increasing cluster size, Sm, Tb and Tm slowly approach their bulk generalized cohesive energies values of respectively 2.64, 2.69 and 2.96~eV~\cite{anna1}. These bulk generalized cohesive energies change only gradually through the whole rare-earth series \cite{borje1}. Translating this behavior also to the clusters, the Sm, Tb and Tm data points can be extrapolated (dashed lines in Fig.~\ref{VALENCERESULTS}) to make predictions for all rare-earth clusters up to a size of 8 atoms. For example, the dimers of Dy, Ho and Er are predicted to be divalent, while for larger cluster sizes they are trivalent. Note however that the behavior of the generalized cohesive energy of the Sm dimer somewhat differs from the rest, i.e. 3-8 atom clusters and the bulk. This discrepancy is the result of a larger 5$d$ electron contribution to the binding for the Sm dimer compared to the Tb and Tm dimer. For clusters larger than 3 atoms this difference in binding becomes negligible small. Thus, the extrapolation for the dimers should be interpreted with a little bit of caution.     

So far only pure divalent and trivalent configurations have been compared. However, Eq.~\ref{BORNMIX} is also used to consider possible mixed valence configurations. As an example the results for Tb$_3$, Tm$_5$, Tm$_6$, Sm$_7$ and Sm$_8$ are shown in Table~\ref{MIXEDTABLE}. Here the latter four are chosen, because they are close to a valence transition. Contrary Tb$_3$ is a simple example, that is far from a valence transition. Note that all (mixed) valence configurations are compared with respect to the pure trivalent configuration. Thus a positive energy difference means that the pure trivalent configuration is more favorable and negative values mean that some or all atoms will be in the divalent configuration.

\begingroup
\squeezetable
\begin{table}[b]
\begin{tabular}{|l|c|c|}
\hline
& & \\[-2.2ex]
\textbf{System} & \textbf{Configuration x} & \textbf{\textit{$E_{\rm triv}-E_{\rm x}$ }}\\
                           &  $(n_{\rm div}-n_{\rm triv})$    & (eV/atom)\\
\hline
& & \\[-2.2ex]
Tb$_3$ & 3-0 & 1.1\\
       & 2-1 & 0.99\\
        &1-2 & 0.57\\
\hline
& & \\[-2.2ex]
Tm$_5$ & 5-0 & -0.1\\
        & 4-1 & 0.06\\
        & 3-2 & 0.06\\
        & 2-3 & 0.03\\
        & 1-4 & 0.02\\
\hline
& & \\[-2.2ex]
Tm$_6$ & 6-0 & 0.09\\
        & 5-1 & 0.19\\
        & 4-2 & 0.16\\
        & 3-3 & 0.17\\
        & 2-4 & 0.14\\
        & 1-5 & 0.07\\
\hline
& & \\[-2.2ex]
Sm$_7$ & 7-0 & -0.13\\
        & 6-1 & 0.12\\
        & 5-2 & 0.16\\
        & 4-3 & 0.08\\
        & 3-4 & 0.05\\
        & 2-5 & 0.11\\
        & 1-6 & 0.16\\
\hline
& & \\[-2.2ex]
Sm$_8$ & 8-0 & -0.02\\
        & 7-1 & 0.07\\
        & 6-2 & 0.14\\
        & 5-3 & 0.09\\
        & 4-4 & 0.24\\
        & 3-5 & -0.01\\
        & 2-6 & 0.03\\
        & 1-7 & 0.06\\
\hline
\end{tabular}
\caption{The energy difference between the pure trivalent $E_{\rm triv}$ and other possible configurations $E_{\rm x}$ is given in column three in eV per atom. Here a positive number means that the pure trivalent configuration is more favorable and negative values mean that some or all atoms will be in the divalent configuration. In the second column the first and second number represent respectively the total number of divalent atoms $n_{\rm div}$ and trivalent atoms $n_{\rm triv}$ of configuration $x$. The first column describes the system.}
\label{MIXEDTABLE}
\end{table}
\endgroup

Table~\ref{MIXEDTABLE} shows that for Tm there is an abrupt change from pure divalent to trivalent, when the cluster size changes from 5 to 6. These pure states are favorable over mixed valence states by roughly less than 0.1~eV/atom. This abrupt valence change appears rather unexpectedly, having in mind that rough surfaces of Tm bulk are divalent~\cite{surface2}. For Sm$_7$ and Sm$_8$ also the pure divalent state is preferred over the mixed valence states by about 0.1 and 0.01~eV/atom respectively. However, the errors involved in these calculations are of the order of 0.1~\cite{borje1,anna1,anna2,borje2}, which makes it difficult to resolve energy differences of this size. Further, at finite temperatures the mixed valence configurations could become more favorable due to their higher entropy. Thus, the absence of mixed valence configurations in Table~\ref{MIXEDTABLE}, cannot be used rigorously to exclude mixed valence states of these clusters, especially when the energy difference between configurations is of order 0.1~eV/atom or smaller. In experiments on rare-earth clusters incorporated in an Ar matrix, abrupt valence changes were indeed observed~\cite{argoncage,argoncage2}. In Ref.~\onlinecite{argoncage2} the valence transition for Sm and Tm clusters is observed at a size of respectively 6 and 10 atoms, which is in agreement with our results. However, the results of Ref.~\onlinecite{argoncage} for Pr, Nd and Sm clusters do not agree with our data. This is likely due to the fact that in the experiments of Refs.~\onlinecite{argoncage,argoncage2} it is very hard to accurately estimate the cluster size.

\subsection{Spectral properties}
\begin{figure}[!hbt]
\includegraphics[trim= 21 110 10 10, clip, width=8cm]{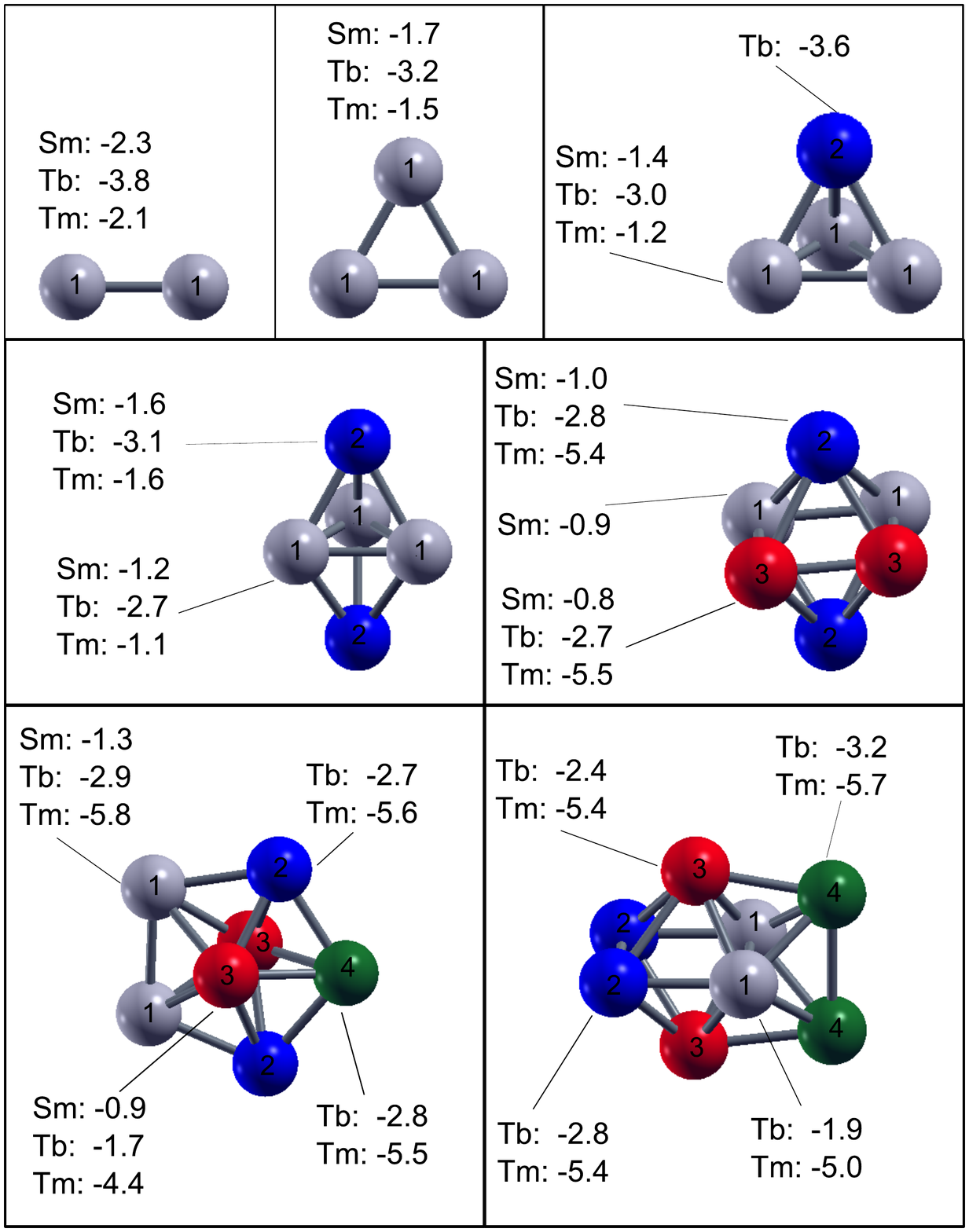}
\caption{Results of the calculated site dependent first 4$f$ peak position below the Fermi level for Sm, Tb and Tm clusters from 2-8 atoms. Equivalent atoms are colored and numbered the same. The number after the rare-earth element corresponds to the first 4$f$ peak position below the Fermi level in eV. For the 4 and 6 atom cluster respectively the atom with number 2 and the atoms with number 1 are only inequivalent for the elements indicated. Thus, for Tb$_{6}$ and Tm$_{6}$ sites 1 and 3 are equivalent, and for Sm$_{7}$ sites 1, 2 and 4 are equivalent.} 
\label{HUBBARDU_4f}
\end{figure}

We now turn our attention to the calculation of spectroscopic information, since it is a very natural way to experimentally detect the valence. To this end we adopt the HIA. However, before the HIA can be used, the number of 4$f$ electrons and the first 4$f$ peak position below the Fermi level should be known for the clusters (see Section~\ref{doc}). The results of the calculations for the first 4$f$ peak position below the Fermi level are presented in Fig.~\ref{HUBBARDU_4f}. Note that for example for the 4 and 6 atom clusters for the blue atom 2 and the grey atoms 1 only one rare-earth element is specified. This means that for the skipped rare-earth elements these are equivalent atoms, i.e. for Sm$_4$ and Tm$_4$ sites 1 and 2 are equivalent, for Tb$_{6}$ and Tm$_{6}$ sites 1 and 3 are equivalent, and for Sm$_{7}$ sites 1, 2 and 4 are equivalent. For the 8-atom cluster information on Sm is not included, because it has a completely different geometry.

\begin{figure}[!ht]
%\begin{center}
%\resizebox{7.5cm}{!} {
\includegraphics[trim=130 1 140 1, clip, width=8cm, scale=0.5]{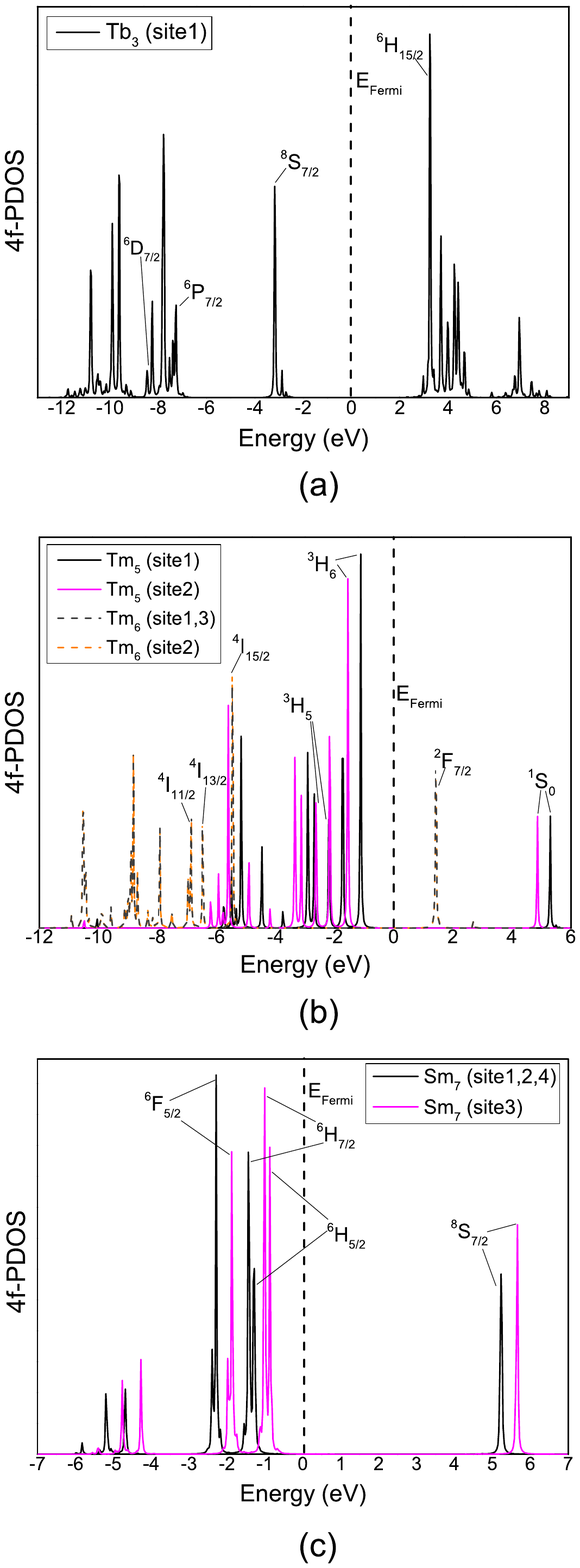}
%}
%\end{center}
\caption{The 4$f$ PDOS calculated with the HIA is plotted for (a) trivalent Tb$_3$, (b) divalent Tm$_5$ and 
trivalent Tm$_6$, and (c) divalent Sm$_7$. Site 1, site 2 etc. refer to the numbered atoms in Fig.~\ref{HUBBARDU_4f}.}
\label{SPECTRA}
\end{figure}
 
The calculations of the first 4$f$ peak position below the Fermi level presented in Fig.~\ref{HUBBARDU_4f} are for the ground state. For Sm and Tb this means respectively a divalent and trivalent configuration up to a cluster size of 8. It can be observed that this peak position is predicted to be quite site dependent. Also, for Tm the binding energy of the first 4$f$ peak below the Fermi level decreases, when the valence transition (at a cluster size of 6 atoms) is approached. This trend is also clear for Sm clusters, as Fig.~\ref{HUBBARDU_4f} shows. Hence, data in this figure, as well as the data in Fig.~\ref{VALENCERESULTS}, suggest that also for Sm a valence transition will occur, for a cluster size larger than 8 atoms. Interestingly, divalent Sm is non magnetic, while trivalent Sm is magnetic according to Hund's rules. For Tb it is clear that the first 4$f$ peak below the Fermi level is approaching the bulk position, which is at 2.2~eV below the Fermi level~\cite{borje2}. Similary the first 4$f$ peak positions below the Fermi level of Tm$_6$-Tm$_8$ are already quite close to that of trivalent Tm bulk (at 4.5~eV below the Fermi level~\cite{borje2}).

%A Hubbard $U$ of 8~eV is used for the calculations of the spectral properties of the Sm, Tb and Tm clusters. This value of $U$ is commonly %used for the 4$f$ shell of the rare-earths~\cite{lars1,impl1,leb1,leb2}. In addition to the Hubbard $U$ parameter also the onsite exchange %interaction $J$ is needed. For the $J$ it is well known that it is almost system independent~\cite{hubj1,hubj2} and is therefore taken to %be 1~eV~\cite{impl1,leb1,leb2}.

Finally, with the valence (Fig.~\ref{VALENCERESULTS}) and first 4$f$ peak position (Fig.~\ref{HUBBARDU_4f}) at hand, HIA calculations are performed. More precisely, the 4$f$ partial density of states (4$f$ PDOS) for Tb$_3$, Tm$_5$ and Tm$_6$, and Sm$_7$ are calculated with the HIA. In Fig~\ref{SPECTRA} the 4$f$ PDOS is presented for (a) trivalent Tb$_3$, (b) divalent Tm$_5$ and trivalent Tm$_6$ and (c) divalent Sm$_7$. Here the site numbers between brackets in the legend refer to the numbered atoms in Fig.~\ref{HUBBARDU_4f} of the corresponding cluster size. For Tb$_3$ all three atoms are equivalent so in Fig.~\ref{SPECTRA}(a) only one site is indicated.  
%\begin{figure*}[!ht]
%\includegraphics[width=1.00\textwidth]{fig3-new.eps}
%\caption{The with HIA calculated 4$f$ PDOS is plotted for (a) trivalent Tb$_3$, (b) divalent Tm$_5$ and 
%trivalent Tm$_6$, and (c) divalent Sm$_7$. Site 1, site 2 etc. refer to the numbered atoms in Fig.~\ref{HUBBARDU_4f}.}
%\label{SPECTRA}
%\end{figure*}
For all plots of Fig.~\ref{SPECTRA} the first 4$f$ peak position below the Fermi level (corresponding to zero energy) is the same as the calculated peak positions of Fig.~\ref{HUBBARDU_4f}, since it is fixed to that value by the definition of the double counting~\cite{impl1}. Also the atomic multiplet structure can be observed clearly in all these plots. For the reader's convenience some of the peaks are indicated. Further, from  Fig.~\ref{SPECTRA} (b) it is clear that the valency strongly affects the spectrum, and this should hence be a clear possibility to experimentally detect the predicted valence stabilities. Also the site dependence of the spectrum can be observed here and in Fig.~\ref{SPECTRA} (c) for Sm$_{7}$. %Note that for Sm$_7$, Fig.~\ref{SPECTRA} (c), site 2 and 4 (see Fig.~\ref{HUBBARDU_4f}) have the same first 4$f$ peak postion although the sites are inequivalent. Due to this equivalent first 4$f$ peak position their spectra look the same, i.e. are on top of each other.

\section{Conclusion and discussion}
In this investigation we outline how the valence stability of rare-earth clusters evolves as a function of cluster size. From first principles theory, combined with the Born-Haber cycle and experimental information of the atomic electronic configuration we show that Sm and Tb clusters are respectively purely divalent and trivalent, respectively, up to a size of 8. Larger clusters of Tb are not expected to have a valence transition, whereas from extrapolation we predict that Sm clusters with 9-10 atoms, or more, will undergo a transition to a trivalent or a mixed valent configuration. As concerns the valence transition of Tm, we find that there is a transition from divalent to trivalent for clusters with six atoms or more. However, the energy difference between different mixed valence configurations is small, as is the energy difference between a mixed valence configuration and an integer valence state. The same holds for Sm$_7$ and in particular Sm$_8$. Sm$_8$ actually has several electronic configurations that are all within 10-100~meV. Unfortunately the accuracy of the valence stability calculations used in this study is approaching these energy differences. Therefore a mixed valence situation for both Tm and Sm clusters may very well be a reality for carefully chosen cluster sizes. This holds even more true at finite temperatures, where mixed valency becomes more favorable due to its larger entropy. An experimental investigation of this prediction would be highly interesting and could potentially also shine light into the finer details of mixed valency. The decisive property of mixed valence systems~\cite{IV1,IV2} is that `fast' physical measurements (with characteristic times faster than, roughly, $10^{-13}$ s) give snapshots corresponding to a random static mixture of divalent and trivalent ions whereas `slow' (or static) measurements give the values average between typical for divalent and trivalent compounds. For example, the atomic volume of mixed valence compounds are intermediate between the values typical for isostructural compounds of divalent and trivalent elements. At the same time, `fast' core-level spectroscopy should give a mixture of the lines (e.g., L3 or L2 spectra) corresponding to Sm$^{2+}$ and Sm$^{3+}$ (or divalent and trivalent Sm), and this seems to be the most convenient experimental way to probe the mixed valence state for the clusters. Note that the difference between homogeneous and inhomogeneous mixed valence crucially important for the bulk \cite{IV1,IV2} is not well-defined for clusters where in general not all atoms have equivalent structural positions, a fact that may be utilized to shed light on finer details of mixed valence.

\begin{acknowledgments}
We acknowledge support from the Swedish Research Council (VR), eSSENCE, STANDUPP, and the Swedish National Allocations Committee (SNIC/SNAC). The Nederlandse Organisatie voor Wetenschappelijk Onderzoek (NWO) and SURFsara are acknowledged for the usage of the LISA supercomputer and their support. The calculations were also performed on resources provided by the Swedish National Infrastructure for Computing (SNIC) at the National Supercomputer Center (NSC), the High Performance Computing Center North (HPC2N) and the Uppsala Multidisciplinary Center for Advanced Computational Science (UPPMAX). O.E. also acknowledges support from ERC (project 247062 - ASD) and the KAW foundation. A.D. acknowledges financial support from KAW and KVA. M.I.K. acknowledges a support by European ResearchCouncil (ERC) Grant No. 338957. B.J. acknowledges the support from the European Research Council (ERC-2008-AdG\_228074).
\end{acknowledgments}

%\bibliography{strings,kylie}

% %%%%%%%%%%%%%%%%%%%%%%%%%%%%%%%%%%%%%%%%%%%%%%%%%%%%%%%%%%%%%%%%%%%%%%%
% %                         BIBLIOGRAPHY
% %%%%%%%%%%%%%%%%%%%%%%%%%%%%%%%%%%%%%%%%%%%%%%%%%%%%%%%%%%%%%%%%%%%%%%%

\end{document}